**Ion Redistribution and Lattice-Distortion–Modulated *d*-Electron Correlation in Layered Multiferroic $ABP_2S_6$ Probed by X-ray Photoelectron Spectroscopy**

Jing Tang[1], Ruiqi Lu[1], J. Anibal Boscoboinik[2], Xi Ling[1,3,4,*]

[1]Division of Materials Science and Engineering, Boston University, Boston, Massachusetts 02215, USA

[2]Center for Functional Nanomaterials, Brookhaven National Laboratory, Upton, New York 11973, USA

[3]Department of Chemistry, Boston University, Boston, Massachusetts 02215, USA

[4]The Photonics Center, Boston University, Boston, Massachusetts 02215, USA

*Email: xiling@bu.edu

**Abstract**

Ferroelectric phase transitions are driven by ion displacement-induced symmetry breaking, yet their chemical signatures are rarely probed directly at the atomic level. Here, we establish temperature-dependent X-ray photoelectron spectroscopy (XPS) as a powerful and quantitative probe of ferroelectric phase transitions by tracking ion migration–induced changes in local chemical environments. Using van der Waals (vdW) layered $ABP_2S_6$ materials as a model system, we systematically compare $CuCrP_2S_6$ (CCPS), which undergoes a ferroelectric phase transition, with $AgCrP_2S_6$ (ACPS), which remains paraelectric over the same temperature range. While ACPS exhibits only smooth trivial thermal broadening and shifts in core-level spectra, CCPS shows pronounced and enormously large temperature-dependent XPS anomalies across the ferroelectric transition, demonstrating that XPS is intrinsically sensitive to ferroelectric order. Detailed analysis of the Cu *2p* spectra reveals that these anomalies originate from temperature-driven $Cu^+$ ion migration associated with polarization switching, whereas concurrent changes in the Cr core levels reflect secondary lattice squeezing and modulation of *d*-electron correlation induced by Cu displacement. Supported by a ferroelectric structural model, we further correlate XPS binding-energy shifts with the magnitude of Cu-ion migration, enabling a quantitative estimation of ionic displacement directly from spectroscopic observables. This work also highlights the rich information provided by temperature-dependent XPS for detecting and quantifying ferroelectric phase transitions and ion migration, offering a general chemical probe for ferroic materials and ion-displacement–driven phase transitions.

Ferroelectric materials are of fundamental and technological importance due to their nonvolatile polarization and potential applications in memory, sensing, and electronic devices[1]. At the microscopic level, ferroelectric phase transitions arise from symmetry breaking accompanied by collective ion migrations, leading to a spontaneous electric polarization[2]. While this phenomenon is often described in terms of macroscopic polarization and crystallographic symmetry changes, its microscopic origin is fundamentally rooted in atomic-scale rearrangements of ions and the associated reconstruction of local chemical environments. Despite extensive studies of ferroelectricity over several decades, how the chemical bonding and local electronic environments of specific ions evolve across a ferroelectric phase transition remains largely unexplored. A variety of experimental techniques have been employed to probe ferroelectric phase transitions, including X-ray and neutron diffraction, electrical polarization measurements[3], and vibrational spectroscopies such as Raman and infrared spectroscopy[4]. Nevertheless, diffraction-based methods provide averaged structural information with limited sensitivity to the associated changes in local electronic and chemical environments. Electrical measurements directly access macroscopic order parameters without identifying its microscopic ionic origins. Vibrational spectroscopies are sensitive to symmetry breaking and lattice dynamics, yet they do not directly resolve element-specific chemical states. Consequently, a critical gap remains, as there is currently no widely established experimental method capable of directly tracking the evolution of the local chemical environment of individual ions across a ferroelectric phase transition.

X-ray photoelectron spectroscopy (XPS) is conventionally regarded as an element-specific spectroscopic technique widely used to determine the elemental composition and chemical states of materials. Fundamentally, however, the binding energy of a core-level electron reflects the local electrostatic potential and bonding environment surrounding the atoms[5]. Ferroelectric phase transitions, which involve ion displacement, polarization-induced internal fields, and lattice distortion, are therefore intrinsically expected to modify core-level binding energies. Although XPS primarily probes the near-surface region, its nanometer-scale information depth enables probing multiple layers of freshly cleaved van der Waals (vdW) materials[6]. Despite this intrinsic sensitivity, XPS has rarely been explored as a spectroscopic probe of ferroelectric phase transitions, and systematic temperature dependent XPS studies across ferroelectric transitions remain notably scarce. Establishing whether XPS can directly detect and quantify ferroelectric ion displacement and change of chemical environment would thus open a new chemical perspective on ferroic phase transitions.

VdW layered $ABP_2S_6$ (A and B are metal cations) materials provide an ideal platform to probe the correlation between XPS signatures and the ferroic phase transition[7]. In these materials, the A-site or B-site cation occupies a weakly bonded, low-coordination environment, making it highly susceptible to displacement and symmetry breaking. Importantly, closely related members in this family exhibit distinct ferroic behaviors. For example, $CuCrP_2S_6$ (CCPS) undergoes a ferroelectric phase transition driven by $Cu^+$ cations ion displacement around 190 K [8], whereas $AgCrP_2S_6$ (ACPS) remains paraelectric over this temperature range while preserving the same structural framework[9,10]. This unique material pair enables a direct comparison that isolates the effects of ferroelectricity from thermal broadening or generic temperature-induced spectral evolution. Here, we report pronounced and nonmonotonic changes in core-level XPS spectra over a temperature range of 175–298 K, encompassing the ferroelectric phase transition in CCPS, while no such behavior is observed in ACPS. These XPS spectral anomalies observed in CCPS are attributed to $Cu^+$ ion redistribution and associated lattice distortions that modify the local chemical environment of Cr. Furthermore, we establish a quantitative correlation between XPS binding-energy shifts and the magnitude of Cu-ion displacement, enabling spectroscopic estimation and visualization of ionic migration across the ferroelectric transition. This work demonstrate temperature-dependent XPS can serve as an element-specific probe of the microscopic ionic rearrangements and coupled lattice distortions underlying ferroic phase transitions.

## Results

### Modeling XPS Responses to Cu-Ion Displacement and Lattice Distortion

To establish a spectroscopic basis for tracking Cu-ion displacement and redistribution among local coordination environments, we first elucidate the spectroscopic mechanism that links local structural rearrangements to core-level spectroscopic responses. Beyond its conventional use for static chemical-state analysis, XPS intrinsically encodes information on local coordination symmetry, electrostatic potential, and lattice strain through core-level binding energies and multiplet electron configurations[11,12]. Here, we focus on two representative spectroscopic responses in CCPS: Cu-ion-displacement-induced core-level binding-energy shifts and Cr-lattice-distortion-induced multiplet reshaping. We use first-principles Δ-self-consistent-field (ΔSCF) core-level calculations[13] to illustrate the Cu ion-migration-induced binding energy shifts in a $CuS_6$ octahedron, and use crystal-field multiplet (CFM) simulations[14,15] for the lattice-distortion–induced multiplet reshaping of Cr ions in a $CrS_6$ octahedron (Figure 1). The $CuS_6$ and $CrS_6$ octahedra serve as simplified models of the local coordination environments of the Cu and Cr ions in the CCPS crystal structure.

To quantify how $Cu^+$ ion displacement affects the Cu $2p_{3/2}$ core-level binding energy, we performed ΔSCF core-level calculations on three representative Cu coordination environments in the $CuS_6$ octahedron (Figures 1a and 1b), generated by displacing $Cu^+$ within the lattice obtained. These configurations differ in their Cu–S bond lengths and local coordination geometry while retaining the same formal $Cu^+$ oxidation state[13]. The site-resolved Cu $2p_{3/2}$ binding energy is evaluated from the ΔSCF total-energy difference between the ground state and a core-ionized final state, represented by a dedicated excited-state pseudopotential carrying one Cu *2p* core hole on the absorber atom, as implemented in Quantum ESPRESSO (see Methods and Supporting Note 1). The site-resolved shift is reported as $\Delta E_{\mathrm{BE}}$ = BE (site i) − BE (site 1), with BE = $E_{\mathrm{ch}} - E_{\mathrm{gs}}$ (site i labels the displacement site i of the $Cu^+$ ion from the reference site 1; $E_{\mathrm{ch}}$ is the core-hole total energy and $E_{\mathrm{gs}}$ is the ground-state total energy). Across the three configurations, the calculated Cu $2p_{3/2}$ binding energy decreases monotonically, reaching $\Delta E_{\mathrm{BE}} \approx -1.4$ eV at site 3 relative to site 1 (Figure 1b). The calculated total energies, Cu displacements, and Cu–S coordination parameters are summarized in Supporting Table S1. This result establishes that $Cu^+$ ion displacement could produce core-level binding-energy shifts without a change in formal oxidation state, providing a quantitative basis for interpreting the sensitivity of Cu *2p* XPS to local coordination changes.

$Cu^+$ displacement further perturbs the surrounding $CrS_6$ framework, in particular, a trigonal squeezing of the Cr-centered $CrS_6$ octahedra (Figures 1c and 1d), which lifts the degeneracy of the Cr *3d* orbitals and modulates the on-site *d*-electron correlation[15,16]. To capture how this lattice squeezing reshapes the Cr *2p* core-level spectrum, we performed CFM simulations for a $Cr^{3+}$ cation in a $CrS_6$ coordination environment using the Crispy software package[16], which solves the multi-electron problem within the Quanty backend and explicitly includes intra-atomic Coulomb and exchange interactions, *2p* and *3d* spin–orbit coupling, and a reduced-symmetry $C_{3v}$ crystal field parameterized by the trigonal distortion strength (*Dτ*)[16,17]. Figure 1d shows the simulated Cr *2p* spectra as a function of *Dτ*. Making *Dτ* more negative progressively reshapes the $2p_{3/2}$ multiplet: the leading peak shifts toward higher binding energy slightly, the high-binding-energy multiplet feature gains spectral weight, and the $2p_{3/2}$–$2p_{1/2}$ separation contracts. The insets of Figure 1d show the corresponding diagonal Cr *3d* crystal-field matrix elements: $a_1$ ($t_{2g}$) rises and crosses above *e* ($e_g$) as *Dτ* becomes more negative. These diagrams illustrate the model basis; the two *e* components can mix, so the plotted diagonal values are not the eigenvalues of the full crystal-field Hamiltonian. The parameter sets used for Figures 1d and 4 are specified in Methods.

Together, the ΔSCF core-level-calculated Cu *2p* binding-energy shifts and the CFM-simulated Cr *2p* multiplet reshaping establish a spectroscopic framework in which both $Cu^+$ ion displacement and the associated $CrS_6$ lattice distortion are directly encoded in the core-level XPS spectra of CCPS. These results provide the theoretical foundation for interpreting the experimentally observed temperature-dependent XPS data presented in the following sections.

**Temperature-Dependent XPS across the Ferroelectric Phase Transition in CCPS**

It is widely reported that temperature drives $Cu^+$ ion displacement and redistribution in CCPS, leading to multiple ferroic states, each characterized by a distinct $Cu^+$ arrangement[18,19]. Figure 2a illustrates the crystal structures of CCPS viewed along the basal plane and perpendicular to the layers. CCPS crystallizes in the monoclinic system with space group *C*2/*c* in the paraelectric phase at room temperature[8,20]. The structure consists of $[P_2S_6]^{4-}$ ethane-like building blocks separated by vdW gaps, with Cu and Cr cations occupying distinct octahedral sites coordinated by six sulfur atoms within each layer. With temperature decreasing, $Cu^+$ ions arrangement changes from occupying the average center position within each vdW layer in a paraelectric state to occupying one-side of the off-center position in ferroelectric (FE) state below 333 K. Between 190 and 333 K, CCPS exhibits stable FE state with out-of-plane symmetry breaking-induced ferroelectric polarization[21,22]. Below 190 K, it transitions into a quasi-antiferroelectric (Q-AFE) state, in which the $Cu^+$ ions are reported to be randomly distributed between the upper and lower sides within each layer[18,22,23]. Replacing $Cu^+$ with $Ag^+$ ions yields ACPS, which preserves the similar structural framework but lacks a ferroelectric phase transition in the investigated temperature range from 175 to 298 K[9,10]. Specifically, $Ag^{+}$ ions in ACPS remain centered within each layer, providing a control system that isolates temperature-dependent spectral changes unrelated to ferroelectricity.

Figure 2b compares the Cu *2p*, Cr *2p*, S *2p*, and P *2p* XPS spectra of CCPS measured at 175 K and 298 K, corresponding to the Q-AFE and FE states, respectively. Pronounced differences in both binding energy and spectral line shape are observed between the two temperatures. The core-level peaks shift toward higher binding energy in the Q-AFE state, while the Cu and Cr *2p* spectra undergo clear line-shape changes and the P and S *2p* spectra exhibit more pronounced high-binding-energy shoulders in the FE state. These changes are consistent with temperature-dependent evolution of the local electronic environments. In contrast, the corresponding Cr *2p*, Ag *3d*, S *2p*, and P *2p* spectra of ACPS at 175 K and 298 K retain essentially unchanged line shapes and exhibit only small binding-energy shifts (Figure *2d*). This direct comparison highlights the pronounced spectral response of CCPS across the transition.

The temperature-dependent core-level heat maps further resolve this contrast. CCPS shows pronounced spectral evolution near 190 K, marked by the white dashed line in Figure 2c, together with a progressive shift of the Cu *2p* peak toward higher binding energy upon cooling above the transition region. In comparison, ACPS shows only weak, smooth spectral evolution over the same temperature range, with no comparable anomaly (Figure 2e). The temperature dependent spectra have been shown in SI Figure S6 and S7. We also extracted the VBM positions from the intersections of linear fitting to the valence-band leading edge and baseline (Supporting Figure S8); the resulting offsets relative to room temperature are shown in the insets of Figures 2c and 2e. The VBM of CCPS exhibits about 0.8 eV, nonmonotonic temperature dependence, including a change across the transition region near 190 K. By contrast, the ACPS VBM shifts only slightly toward lower binding energy, by approximately 0.15 eV, without a comparable transition-like feature. The concurrent changes in the CCPS core-level spectra and VBM support an electronic-structure response coupled to the transition-associated rearrangement of $Cu^{+}$ ions and their local coordination environments. Earlier electrical measurements identified $Cu^{+}$ ion hopping as a mechanism of ionic conduction in CCPS[20], providing a link between $Cu^{+}$ ion dynamics and charge transport. The temperature-dependent energy shifts of the Cu, Cr, S, and P core levels and the VBM, each referenced to its room-temperature value, are compared in Supporting Figure S9. In addition to the overall energy shifts, the Cu *2p* spectrum develops two components near 206 K, with their separation evolving upon further cooling. This component evolution cannot be accounted for by a uniform energy shift alone and provides evidence for distinct local Cu environments. As an additional control experiment, we measure the temperature-dependent XPS spectra of $CuInP_2S_6$ (CIPS) over the same temperature range, where CIPS undergoes no phase transition[24] (Supporting Note 5). Accordingly, its temperature-dependent XPS heat maps show no discontinuity in any core level (Supporting Figures S10 and S11). The contrasting line-shape evolution in CCPS, ACPS, and CIPS supports an association between the CCPS anomalies and ferroic $Cu^+$ redistribution. It is also worth noting that all spectra are acquired under rigorously controlled charge-neutralization conditions to suppress sample charging, excluding the contribution of Fermi level shift to the observed core-level spectra changes across the measured temperature range.

## Cu *2p* Spectral Evolution Revealing an Intermediate State of Ferroelectric Ion Redistribution

To further elucidate the microscopic origin of the temperature-dependent XPS anomalies in CCPS, we investigate the fine evolution of the Cu $2p_{3/2}$ core-level spectra across the temperature range from 175 K to 298 K (Fig. 3a). At room temperature, a single Cu $2p_{3/2}$ peak is observed at 931.2 eV, consistent with the model shown in the top schematic of Fig. 3a, in which $Cu^+$ occupies an upper off-center site within each $CuS_6$ coordination environment. As temperature decreases, the peak undergoes pronounced broadening and splits into two well-resolved components near 206 K. Upon further cooling to 175 K, the peak separation grows and the spectral weight progressively transfers from the higher BE component to the lower BE component. We denote the higher BE component as $p_t$ and the lower BE component as $p_b$, corresponding to $Cu^+$ ion at the top ($Cu_t$) and bottom ($Cu_b$) off-centered positions within each vdW layer, as illustrated in the bottom schematic of Fig. 3a. The middle schematic represents the transient state in which $Cu^+$ ion explores intermediate positions before settling.

The emergence of two components of the Cu $2p_{3/2}$ peak is consistent with distinct local Cu environments becoming spectroscopically distinguishable across the transition region. Within the structural model considered here, the two Cu sites have different local Cu–S coordination environments, which can give rise to distinct Cu $2p_{3/2}$ binding energies [25]. The contributions of the local electrostatic potential and electronic relaxation following core-hole creation are examined in the first-principles analysis below. The higher-BE component $p_t$ is accordingly assigned to $Cu_t$ and the lower-BE component $p_b$ to $Cu_b$. A detailed first-principles analysis of the microscopic origin of the splitting, including the role of Cu–S bond-length asymmetry, multi-well off-centering, and configurational variation, is presented in the later section. In this section, we treat the redistribution using an effective two-environment model, between $Cu_t$ and $Cu_b$.

Figure 3b shows the fitted BE positions of the high-temperature Cu $2p_{3/2}$ peak and the two lower-temperature components as a function of temperature. The single $p_{Cu}$ component (in red) is observed between 220 K and 300 K at 931.2–931.5 eV. Below 206 K, $p_{Cu}$ disappears and is replaced by two components: $p_t$ (in green) around 932.2 eV and $p_b$ (in blue) around 931.4 eV, corresponding to a low-temperature splitting of around 0.9-1.0 eV. The gray shaded window from 175 to 200 K marks the FE to Q-AFE crossover, within which $p_t$ and $p_b$ coexist and their relative spectral weights evolve continuously with temperature.

Within this two-environment model, their integrated peak intensities $I_{p_t}$ and $I_{p_b}$ for the $p_t$ and $p_b$ components of Cu $2p_{3/2}$ peaks, respectively provide estimates of the relative populations of $Cu^+$ ion at the two sites. We quantify the extent of $Cu^+$ ion migration by the population fraction $f(Cu_b) = \frac{I_{p_b}}{I_{p_t}+I_{p_b}}$, shown as a function of temperature in Fig. 3c. $f(Cu_b)$ rises continuously from 0.30 at 195 K to 0.45 at 175 K, indicating that $Cu^+$ ion redistribution proceeds gradually rather than through a sudden collective displacement. We fit the trend with a two-state Boltzmann sigmoid $f(Cu_b) = 0.3 + \frac{0.2}{1+e^{\frac{T-T_0}{k}}}$, where $T_c$ is the inflection temperature and $k$ sets the transition width. The best fit yields $T_0$ = 180.0 K and $k$ = 4.5 K. Extrapolation predicts $f(Cu_b)$ reach to 0.5 below 160 K, consistent with the onset of an AFE phase in which $Cu^+$ populates the top and bottom off-centered positions with equal occupancy in an alternating configuration. An independently measured data point at 146 K (in orange circle in Fig. 3c) gives $f(Cu_b) = 0.505$, in quantitative agreement with the projected 1:1 ratio and with the reported Q-AFE-to-AFE transition near 145 K[22,26] (Supporting Note 6). While the Q-AFE phase has been proposed in previous work, its microscopic configuration has remained unclear[8,21,27]. Our results provide a direct spectroscopic visualization of the $Cu^+$ ion migration that defines the Q-AFE state.

To highlight the capability of XPS in probing $Cu^+$ ion redistribution in the CCPS lattice, we use a conventional structural probe, single-crystal XRD to characterize CCPS and ACPS across a wider temperature range from 100 to 300 K (Fig. 3d, 3e). Figure 3d shows the in-plane lattice parameters *a*, *b*, *c* of CCPS extracted from sequential refinement along both cooling (dark) and warming (light) branches. Outside the transition window the three lattice parameters contract smoothly with decreasing temperature. Within the FE to Q-AFE transition window (i.e. 145–190 K, gray shaded), however, all three parameters exhibit an anomalous apparent expansion of 50–100%. Clearly distinct cooling and warming branches are observed, indicating a thermal hysteresis of around 10–15 K, consistent with the first-order character of the transition reported in calorimetric studies[22]. It is worth noting that the apparent expansion is not a real volume increase but reflects the breakdown of a single, well-defined unit cell. Specifically, in the Q-AFE intermediate state, partial $Cu^+$ ion occupancy of inequivalent top and bottom sites destroys translational equivalence, and the average lattice constants returned by single-cell refinement become ill-defined. Single-crystal XRD refinements reveal substantially poorer agreement between the observed and calculated structure factors at 160 K than at 100 and 291 K (Supporting Note 7 and Figure S13). The corresponding $R_1$ values are 0.5544, 0.1249, and 0.0537, respectively, and the tested model at 160 K fails to converge. In contrast, Fig. 3e shows that the lattice parameters of ACPS in which $Ag^+$ remains nearly unchanged across the entire 100–300 K range, with no discernible anomaly. This stark contrast confirms that the CCPS feature is intrinsic to the ferroelectric Cu redistribution rather than to a thermal artifact[10].

As XRD reports the spatially averaged structure, it is therefore blind to the site-specific Cu redistribution that defines the Q-AFE state where the two Cu sublattices are partially populated and no single repeatable unit cell describes the crystal. In contrast, XPS resolves the environment difference between the two $Cu^+$ ions directly through the $p_t/p_b$ splitting and quantifies their relative populations through $f(Cu_b)$. This result highlights that temperature-dependent core-level spectroscopy provides element-specific, statistically resolved access to the ferroelectric $Cu^+$ ion migration that XRD cannot reach.

**Local Distortion–Driven *d*-Electron Correlation Revealed by Cr *2p* XPS**

Beyond the Cu sublattice, the redistribution of $Cu^+$ leaves its imprint on the surrounding $CrS_6$ framework. Because $Cr^{3+}$ carries a partially filled $3d^3$ shell, its $2p$ line shape reports on the local crystal-field symmetry rather than on the formal oxidation state[8,16,28], and we therefore use the temperature-dependent Cr $2p$ spectra, analyzed together with CFM simulations, to read out the local lattice distortion that accompanies Cu redistribution across $T_C$.

The CFM simulations parametrize the symmetry lowering from $O_h$ to $C_{3v}$ through a single trigonal distortion parameter $D\tau$, with all other parameters held fixed (see Methods)[16,17]. Figure 4a shows simulated Cr $2p$ spectra over a series of $D\tau$ values from −0.259 to −0.270 eV; as $D\tau$ becomes more negative when the trigonal distortion grows, the $2p_{3/2}$ leading edge sharpens, the satellite at the high-BE side gains spectral weight, and the overall $2p_{3/2}$–$2p_{1/2}$ separation contracts. Figures 4b–d compare these simulations (in red curves) with experimental Cr $2p$ spectra (in black circles) measured at three representative temperatures spanning the phase transition. At 289.4 K (i.e. FE phase), the experimental Cr $2p_{3/2}$ envelope exhibits a clearly resolved doublet, a leading peak near 574 eV and a second component on the low-BE side near 572 eV (Fig. 4b). The CFM simulation with $D\tau = -0.260$ captures the overall multiplet envelope, consistent with the interpretation that the $CrS_6$ octahedra are already trigonally distorted in the room-temperature FE phase. At 195.6 K (i.e. within the FE to Q-AFE transition window), the $2p_{3/2}$ doublet narrows and the low-BE component weakens into a right-hand shoulder on the leading peak (Fig. 4c). The CFM simulation with $D\tau = -0.267$ captures the overall envelope together with the gradual sharpening of the leading peak and the strengthening of the high-BE satellite. At 175.1 K (i.e. Q-AFE phase), the $2p_{3/2}$ envelope collapses into a nearly single, sharp peak with no resolved low-binding-energy shoulder (Fig. 4d). The CFM simulation with $D\tau = -0.270$ reproduces this near-single-peak line shape together with the maximally sharpened satellite structure. The qualitative agreement between the experimental spectra and the CFM simulations suggests that the temperature-driven evolution of the Cr line shape can be primarily attributed to lattice-distortion-induced changes in the local crystal-field and multiplet structure[16,17]. Within this effective model, the monotonic change in $D\tau$ with decreasing temperature suggests that the $CrS_6$ trigonal squeezing grows continuously as $Cu^+$ ion migration progresses, providing a model-dependent spectroscopic description of the lattice distortion that accompanies the ferroelectric transition.

The CFM simulation also describes the diagonal crystal-field matrix elements of the Cr *3d* shell and their changes with the trigonal distortion. Under the $C_{3v}$ crystal field, the parent $t_{2g}$ manifold of $O_h$ symmetry splits into a non-degenerate $a_1(t_2g)$ component and a doubly degenerate $e(t_2g)$ component, while the $e_g$ manifold transforms as $e(e_g)$. The two e components can mix; the values in Fig. 4e–h are diagonal matrix elements in this symmetry-adapted basis, not the eigenvalues of the full crystal-field Hamiltonian. At $D\tau = -0.260$ (FE phase), the three diagonal elements are $e(t_{2g}) = -0.253$ eV, $a_1(t_{2g}) = +0.280$ eV, and $e(e_g) = +0.113$ eV (Fig. 4e). As $D\tau$ decreases to −0.267 (Fig. 4f) and −0.270 (Fig. 4g), the $a_1(t_2g)$ diagonal element rises monotonically to +0.322 and +0.340 eV, respectively, moving away from the $e(e_g)$ diagonal element, which decreases to +0.097 and +0.090 eV. The $e(t_2g)$ diagonal element changes only slightly, to −0.258 and −0.260 eV, respectively. Figure 4h compiles these diagonal matrix elements as a function of $D\tau$, where all three diagonal matrix elements vary linearly with $D\tau$, with $a_1(t_2g)$ showing by far the strongest dependence, increasing as $D\tau$ becomes more negative, while $e(e_g)$ decreases more slowly and $e(t_2g)$ remains nearly constant. The three sets of color-coded markers in Fig. 4h (i.e. purple, green and yellow) correspond to the three $D\tau$ values shown in Fig. 4e–g (i.e. $D\tau = -0.260$, −0.267, and −0.270, respectively). Thus, the crystal-field-driven redistribution of the *3d* manifold alters the multiplet-state composition of the $3d^3$ ground state and of the $2p^5 3d^3$ final state, producing the systematic Cr *2p* line-shape changes observed in Fig. 4a–d[15,16]. Together, these results support Cr *2p* XPS as an element-specific probe of changes in the local crystal field associated with $Cu^+$ redistribution.

**Understanding of the Microscopic Origin of the Cu *2p* Splitting**

To clarify the microscopic origin of the ~ 0.9-1.0 eV splitting of Cu $2p_{3/2}$ observed below $T_C$, we now turn to first-principles calculations. While the inequivalent Cu sites in the low-temperature Pc structural model provide a qualitative basis for splitting as explained in Figure 1b, this model does not specify which Cu configurations within the underlying multi-well potential are required to reproduce the observed splitting magnitude. We address this question using ΔSCF core-level calculations performed on CCPS with a series of Cu sublattice configurations (Supporting Note 8).

We first applied the ΔSCF core-level calculation to the experimentally refined low-temperature (Pc) structure of CCPS reported previously[8]. This static structure yields a Cu site splitting of $\Delta E_{Cu}$

= +0.25 eV, with $Cu_t$ at higher binding energy than $Cu_b$. This calculated splitting is substantially smaller than the experimentally observed value of 0.9-1.0 eV, indicating that the diffraction-averaged Cu positions alone cannot account for the full spectral splitting. This discrepancy motivates a closer examination of Cu sublattice configurations within the underlying multi-well potential.

The 0.25 eV initial-state contribution to the splitting reflects the difference in the local electrostatic potential at the two Cu sites, as visualized by the deformation charge density $\Delta\rho = \rho_{SCF} - \Sigma\rho_{atom}$ in Fig. 5a,b ($\rho_{SCF}$: self-consistent crystal density; $\rho_{atom}$: superposed isolated-atom densities). The depletion–accumulation pattern around each Cu and its three nearest S neighbors is the characteristic of a polar covalent Cu–S bond. In the refined Pc cell, Cut lies closer to the triangular plane formed by the three upper S atoms and exhibits Cu–S bond lengths of 2.13, 2.17 and 2.20 Å (mean 2.17 Å), whereas $Cu_b$ sits closer to the median (Cr/P) cation plane at the centre of the layer and shows longer bonds of 2.21, 2.22 and 2.29 Å (mean 2.24 Å), giving a difference in mean Cu–S bond length between the two sites of $\delta r \approx 0.07$ Å. One-dimensional $\Delta\rho$ profile along the shortest Cu–S bond (Fig. 5c) gives a bonding-charge maximum of $\Delta\rho_{max} \approx 0.024$ e/bohr3 at 1.27 Å from Cut versus 0.021 e/bohr$^3$ at 1.37 Å from $Cu_b$. On the shorter Cut–S bond the bonding-charge maximum is displaced from the bond midpoint toward S by ≈ 0.2 Å, reflecting the polar covalent character of the bond. This stronger Cut–S covalency is what makes $Cu_t$ the higher-binding-energy site, through two effects that act in the same direction. The splitting therefore receives two contributions of the same sign. The initial-state term originates in the ground-state potential: the shorter, more covalent $Cu_t$–S bonds deepen the electrostatic potential at the $Cu_t$ nucleus and bind its 2p electrons more strongly. The final-state term originates in the response to the core hole: the more localized bonding charge around $Cu_t$ screens the *2p* hole less efficiently, adding a further positive shift. Because both channels act in the same direction, these calculations provide a microscopic, first-principles basis for assigning the higher-BE component $p_t$ to $Cu_t$, in agreement with the symmetry argument given above.

One remaining question is the substantial discrepancy between the DFT-predicted Cu $2p_{3/2}$ energy splitting (i.e. 0.25 eV) and the experimentally observed value (i.e. 0.9-1.0 eV). We recall that the averaged diffraction-refined Cu positions reported for the low-temperature Pc structure used in the calculation represent the spatially and temporally averaged Cu sublattice resolved by XRD, whereas the $Cu^+$ cations in CCPS, and in the broader $ABP_2S_6$ family, are well established to occupy a shallow multi-well potential along the polar *c* axis[7,8]. Because the photoelectron emission process is much faster than the Cu ion migration, XPS probes instantaneous Cu configurations rather than the diffraction-averaged centroid. The measured splitting therefore reflects the ensemble of off-centered Cu sites that are instantaneously populated, rather than the single diffraction-refined reference structure. In particular, when $Cu_t$ migrates from the top site to the bottom site of a given layer, it does not necessarily land at the symmetric-mirror position of the remaining $Cu_t$. Instead, it can populate other off-centred sites within the lower half of the multi-well potential, each of which has a distinct local chemical environment and therefore a distinct Cu $2p_{3/2}$ binding energy.

We further investigate the dependence of the Cu $2p_{3/2}$ splitting energy on the occupied sites of Cu ions.  how We perform two complementary rigid migration scans of the diffraction-refined low-temperature Pc cell along the polar *c* axis and calculate the corresponding ΔBE (Fig. 5d). In the first scenario, $Cu_t$ and $Cu_b$ are migrate symmetrically together toward the central Cr plane by equal amounts $\Delta z$. In this case, ΔBE attains a shallow maximum of ~0.27 eV near $\Delta z \approx 0.3$ Å and then decreases monotonically with further $\Delta z$, eventually changing sign as the two Cu cations approach each other across the Cr plane. Thus, the result shows that equal-displacement configurations is difficult to reach an energy splitting beyond 0.27 eV through this symmetric, mirror-preserving migration. In the second scenario, $Cu_t$ is held at its Maisonneuve position while only $Cu_b$ is displaced. In this case, $Cu_b$ is allowed to migrate to bottom off-centered sites that are not the symmetric mirror of $Cu_t$. In this scenario, ΔBE rises monotonically with Δz and reaches 0.89 eV at $\Delta z \approx 0.96$ Å, in quantitative agreement with the experimental ~0.8 eV. The asymmetric migration thus provides insights into the microscopic Cu sublattice arrangement that reproduces the experimentally observed energy splitting magnitude.

## Conclusion

In conclusion, we establish an atomic-level picture of the ferroelectric phase transition in $CuCrP_2S_6$ through a comprehensive understanding of the temperature-dependent core-level XPS. We observe pronounced Cu $2p_{3/2}$ splitting and spectral-weight evolution over the temperature range spanning the ferroic FE-to-Q-AFE phase transition near 190 K, which is attributed to the inequivalent chemical environment of $Cu^+$ ions during the phase transition. By analyzing the spectra weights of the two components from the Cu *2p3/2* peak splitting, we reveal a continuous redistribution of $Cu^+$ ion between two inequivalent sublattice sites and projecting equipartition ($f_{Cu_b}$ = 0.5) below ~160 K, in line with the reported Q-AFE-to-AFE transition at 145 K. Moreover, the Cr $2p_{3/2}$ peak shape tracks a $Cu^+$-driven trigonal squeezing of the $CrS_6$ octahedra that raises the $a_1$ ($t_{2g}$) diagonal crystal-field matrix element at fixed $Cr^{3+}$ valence.  First-principles ΔSCF calculations provide a model for the origin of the approximately 0.9-1.0 eV Cu $2p_{3/2}$ splitting, providing a microscopic basis for relating the spectral components to inequivalent Cu coordination environments.. Beyond revealing the atomic-scale evolution of ion migration and lattice distortion during the ferroic phase transition, this work also establishes temperature-dependent XPS as a chemically resolved probe of ferroic phase transitions and configurational disorder, providing a framework applicable to other multiferroics and ion-migration-driven phase transitions.

**Methods**

Sample Preparation

Single crystals of $CuCrP_2S_6$ (CCPS), $AgCrP_2S_6$ (ACPS) and $CuInP_2S_6$ (CIPS) were synthesized via the chemical vapor transport (CVT) method[10,27]. High-purity elemental precursors of Cu (or Ag), Cr (or In), P, and S were weighed according to the stoichiometric ratio and thoroughly mixed inside an argon-filled glovebox to avoid oxidation and moisture contamination. The mixed powders were loaded into a quartz ampoule together with a small amount of transport agent (i.e. iodine), evacuated to ~$10^{-5}$ Torr, and flame sealed.

The sealed ampoule was placed in a two-zone horizontal tube furnace with a temperature gradient maintained between the source and growth zones. The source zone was heated to 800 °C, while the growth zone was maintained at a lower temperature of 700 °C, and the reaction was sustained for 14 days to allow crystal growth via vapor-phase transport. After completion of the growth process, the furnace was slowly cooled to room temperature at a controlled rate to minimize thermal stress. Plate-like single crystals were extracted from the growth zone.  SEM and EDS measurements of CCPS and ACPS, along with room-temperature Raman spectroscopy characterization of CCPS are summarized in Supporting Note 2 (Figures S1 and S2).

X-ray Photoelectron Spectroscopy (XPS)

X-ray Photoelectron Spectroscopy (XPS) measurements were performed using a PHI Genesis system equipped with monochromatic Al Kα radiation (1486.6 eV). The system provides high sensitivity and energy resolution for precise determination of elemental composition and chemical states. All spectra were acquired under ultra-high vacuum conditions ($\sim 3 \times 10^{-7}$ Pa) to minimize surface contamination. Core-level spectra were collected with appropriate pass energy settings to ensure reliable peak resolution. Temperature-dependent XPS measurements were conducted using a liquid-nitrogen-cooled sample platen, allowing controlled cooling from room temperature down to 175 K. Samples were allowed to thermally stabilize at each temperature prior to spectral acquisition. The temperature was monitored continuously throughout the measurement to ensure stability. Freshly cleaved crystal surfaces were used to minimize surface oxidation and contamination effects. The bulk specimens were approximately 0.5–1 mm thick and were mechanically held by a copper mask without adhesive, leaving the analysis surface exposed. The analysis area was approximately 50–100 μm and was contained within a continuous crystal surface. Charge compensation used the integrated low-energy electron and $Ar^+$ ion gun dual-neutralization system. Valence-band spectra are presented on the original binding-energy axis without post-acquisition alignment (Supporting Note 4 and Figure S8). VBM offsets are extracted by linear leading-edge and baseline intersection; the core-level position definitions, relative-energy references, and uncertainties are specified in Supporting Note 4. Spectra in Figure 3 and the Cu $2p_{2/3}$ spectrum at were fitted after Shirley backgrounds removing. Area fractions are interpreted within this fitting model. The fittings of S and P peaks in Figure S9 use Shirley backgrounds and constrained spin–orbit doublets, while Cr is represented by the fitted main-envelope maximum. The XPS data at 146 K were collected at the Center for Functional Nanomaterials (CFN), Brookhaven National Laboratory, using a lab-based SPECS ambient-pressure XPS system equipped with a SPECS Phoibos analyzer and a monochromated Al Kα photon source (1486.6 eV). The measurement was carried out under UHV conditions with the sample cooled on a cryogenic stage and allowed to thermally equilibrate at 146 K before acquisition.

X-ray Diffraction (XRD)

Room-temperature X-ray diffraction (XRD) measurements were carried out using a Bruker D8 Discover diffractometer in parallel-beam geometry. The instrument is equipped with a sealed-tube Cu X-ray source operating at 1600 W, a Göbel mirror monochromator, a 0.6 mm primary divergence slit, 2.5° primary and secondary axial Soller slits, and a Venture 1D detector. Diffraction patterns were collected for 1 hour to ensure sufficient counting statistics. For temperature-dependent measurements, the diffractometer was integrated with a Bruker temperature controller, enabling controlled cooling from room temperature down to 100 K. Samples were stabilized at each target temperature prior to data acquisition to ensure thermal equilibrium. All temperature-dependent scans were performed under identical geometric and instrumental conditions to allow reliable comparison of lattice parameter evolution. The CCPS datasets at 100, 160, and 291 K were refined with olex2.refine 1.5 using fixed recorded unit-cell parameters. The tested models, statistics, and structure-factor comparisons are reported in Supporting Note 7 (Figure S13 and Tables S7–S8).

Crystal-Field Multiplet (CFM) Simulations

Cr *2p* XPS spectra were calculated with the crystal-field multiplet approach using the Quanty code[17] via the Crispy interface[16]. The calculation used fixed $2p^6 3d^3$ initial and $2p^5 3d^3$ final configurations. The $Cr^{3+}$ ($3d^3$) ion was modeled in $C_{3v}$ symmetry, including the atomic *3d–3d* and *2p–3d* Coulomb and exchange interactions, the *2p* and *3d* spin–orbit couplings, and the trigonal crystal field parametrized by 10Dq, Dσ and Dτ. Here 10Dq is the cubic crystal-field splitting between the $e_g$ and $t_{2g}$ sets of the parent $O_h$ field, while Dσ and Dτ describe the distortion along the $C_3$ axis and arise from the second- and fourth-order terms of the trigonal potential, respectively. Lowering the symmetry from $O_h$ to $C_{3v}$ splits the $t_{2g}$ manifold into $a_1(t_{2g})$ and $e(t_{2g})$ levels, with $e_g$ becoming $e(e_g)$; the magnitude of this splitting is set by Dσ and Dτ, which are treated here as effective crystal-field parameters. The Slater integrals were scaled to 0.54 and 0.81 for $F^2$ and $F^4$ (*3d*, *3d*) and to 0.80, 0.80 and 0.90 for $F^2$, $G^1$ and $G^3$(*2p*, *3d*), respectively, while the spin–orbit parameters were kept at their atomic values. For Figure 4, the cubic field and Dσ were held fixed (10Dq = 1.20 eV, Dσ = 0.400 eV), and the trigonal parameter Dτ was varied to describe the temperature-associated spectral evolution (Dτ = −0.260, −0.267 and −0.270 eV for experimental temperatures of 289.4, 195.6 and 175.1 K, respectively); Dσ and Dτ are treated as effective crystal-field parameters. The same crystal-field parameters were used for the initial and final configurations. Initial-state Boltzmann populations were evaluated at a fixed calculation temperature of 298 K for all spectra; only Dτ was varied to represent the experimental temperature dependence. The calculated spectra were broadened with a Lorentzian of 0.4 eV (FWHM, core-hole lifetime) and a Gaussian of 0.7 eV (FWHM, instrumental resolution), and were then aligned to the experimental binding-energy scale and normalized for comparison. Figure 1d uses the same 10Dq = 1.20 eV and Dσ = 0.400 eV with illustrative Dτ values of −0.225, −0.268, and −0.315 eV. These values illustrate spectral sensitivity and are not a calibration of geometric compression. The diagrams in Figures 1d and 4e–h show diagonal crystal-field matrix elements rather than the eigenvalues after mixing the two *e* components.

Density-Functional Theory (DFT) Calculations

First-principles calculations of the Cu $2p_{3/2}$ core-level binding energies were performed within the ΔSCF approach as implemented in Quantum ESPRESSO 7.2[29]. Exchange–correlation was treated at the PBE+U level with a Hubbard $U$ = 4.0 eV applied to the Cr *3d* orbitals in the ortho-atomic projection, together with a Grimme-D3 vdW correction to capture inter-layer dispersion. The plane-wave kinetic-energy cutoff was 70 Ry for the wavefunctions and 700 Ry for the charge density, and the Brillouin zone was sampled with a 4 × 2 × 2 Γ-centred Monkhorst–Pack mesh. Cu was described by a PAW pseudopotential; S, Cr and P were described by ultrasoft pseudopotentials, all consistent with PBE. For each absorber Cu site, a dedicated excited-state Cu pseudopotential carrying one *2p* core hole was generated with the *ld1.x* utility using the same all-electron recipe as the ground-state Cu dataset (the valence electron count was increased from 11 to 12 to maintain charge neutrality). The site-resolved Cu *2p* binding energy at site α was extracted as BE(α) = $E_{\mathrm{ch}}(\alpha) - E_{\mathrm{gs}}$, where $E_{\mathrm{ch}}(\alpha)$ is the total SCF energy of the cell with the core hole on site α and $E_{\mathrm{gs}}$ is the ground-state total SCF energy of the same cell; only relative shifts (site-to-site differences) are reported in the main text, since the absolute BE depends on the pseudopotential reference and is not directly observable.

Two complementary sets of ΔSCF calculations were performed. (i) For Figure 1, three representative Cu coordination environments arising from symmetry-lowering $Cu^{+}$ displacements within the layered CCPS lattice were taken as input geometries (sites 1, 2 and 3, ordered by increasing off-centering magnitude). BE(α) was evaluated at each of these three sites and reported relative to BE (site 1) to give the site-to-site Cu *2p* shift. (ii) For Figure 5, the experimentally refined low-temperature Pc structure of CCPS reported by Maisonneuve *et al.*[8] was used as the input geometry. ΔSCF was applied to the two inequivalent Cu sites $Cu_t$ and $Cu_b$ to obtain the splitting $\Delta E_{Cu} = BE(Cu_t) - BE(Cu_b)$. Two rigid *c*-axis displacement scans were then carried out on the same Maisonneuve Pc cell to probe the sensitivity of $\Delta E_{Cu}$ to the Cu off-centering configuration: a symmetric scan in which $Cu_t$ and $Cu_b$ were translated by equal amounts $\Delta z$ toward the central Cr plane (preserving the t/b mirror correspondence), and an asymmetric scan in which only $Cu_b$ was translated while $Cu_t$ was held at its original position; all other atoms were kept fixed in both scans. The deformation charge density shown in Figure 5a–c was constructed as $\Delta\rho = \rho_{SCF} - \Sigma\, \rho_{atom}$, where $\rho_{atom}$ is the spherically averaged self-consistent density of an isolated neutral atom placed at the Pc lattice site.

**References**

1. Scott, J. F. Applications of Modern Ferroelectrics. *Science* **315**, 954–959 (2007).
2. Cohen, R. E. Origin of ferroelectricity in perovskite oxides. *Nature* **358**, 136–138 (1992).
3. Jones, J. L. The use of diffraction in the characterization of piezoelectric materials. *Journal of Electroceramics* **19**, 69–81 (2007).
4. Cummins, H. Z. Brillouin scattering spectroscopy of ferroelectric and ferroelastic phase transitions. *Philosophical Transactions of the Royal Society of London. Series A, Mathematical and Physical Sciences* **293**, 393–405 (1979).
5. Briggs, D. & Grant, J. T. *Surface Analysis by Auger and X-Ray Photoelectron Spectroscopy*. (IM Publications, Chichester, UK, 2003).
6. Tanuma, S., Powell, C. J. & Penn, D. R. Calculation of electron inelastic mean free paths (IMFPs) VII. Reliability of the TPP-2M IMFP predictive equation. *Surface and Interface Analysis* **35**, 268–275 (2003).

7. Susner, M. A., Chyasnavichyus, M., McGuire, M. A., Ganesh, P. & Maksymovych, P. Metal Thio- and Selenophosphates as Multifunctional van der Waals Layered Materials. *Advanced Materials* **29**, 1602852 (2017).

8. Maisonneuve, V., Cajipe, V. B. & Payen, C. Low-temperature neutron powder diffraction study of copper chromium thiophosphate (CuCrP2S6): observation of an ordered, antipolar copper sublattice. *Chemistry of Materials* **5**, 758–760 (1993).

9. Dziaugys, A., Banys, J., Macutkevic, J. & Vysochanskii, Y. Dielectric and electrical properties of AgCrP2S6 and Cu0.2Ag0.8CrP2S6 layered crystals. *Ferroelectrics* **515**, 5–12 (2017).

10. Pawbake, A. *et al.* Crystal Growth of the Quasi-2D Quarternary Compound AgCrP2S6 by Chemical Vapor Transport. *Crystals* **11**, 500 (2021).

11. Egelhoff, W. F. Core-level binding-energy shifts at surfaces and in solids. *Surface Science Reports* **6**, 253–415 (1987).

12. Mason, M. G. Electronic structure of supported small metal clusters. *Physical Review B* **27**, 748–762 (1983).

13. Cavigliasso, G. & Chong, D. P. Accurate density-functional calculation of core-electron binding energies by a total-energy difference approach. *Journal of Chemical Physics* vol. 111 9485–9492 (1999).

14. de Groot, F. M. F. *et al.* 2p x-ray absorption spectroscopy of 3d transition metal systems. *Journal of Electron Spectroscopy and Related Phenomena* **249**, 147061 (2021).

15. Stavitski, E. & de Groot, F. M. F. The CTM4XAS program for EELS and XAS spectral shape analysis of transition metal L edges. *Micron* **41**, 687–694 (2010).

16. Retegan, M. Crispy. https://doi.org/10.5281/zenodo.1008184 (2024).

17. Haverkort, M. W., Zwierzycki, M. & Andersen, O. K. Multiplet ligand-field theory using Wannier orbitals. *Phys. Rev. B* **85**, 165113 (2012).

18. Cho, K. *et al.* Tunable Ferroelectricity in Van der Waals Layered Antiferroelectric CuCrP2S6. *Advanced Functional Materials* **32**, 2204214 (2022).

19. Lai, Y., Song, Z. & Yang, J. High-performance van der Waals antiferroelectric CuCrP2S6-based memristors. *Nature Communications* **14**, 7785 (2023).

20. Maisonneuve, V. *et al.* Ionic conductivity in ferroic CuInP2S6 and CuCrP2S6. *Ferroelectrics* **196**, 257–260 (1997).

21. Lai, Y. *et al.* Direct observation of intrinsic room-temperature ferroelectricity in 2D layered CuCrP2S6. *Nature Communications* **14**, 7359 (2023).

22. Moriya, K. *et al.* Low-temperature calorimetric study of phase transitions in CuCrP2S6. *Solid State Communications* **136**, 173–176 (2005).

23. Bourdon, X., Maisonneuve, V., Cajipe, V. B., Payen, C. & Fischer, J. E. Copper sublattice ordering in layered CuMP2Se6 (M = In, Cr). *Journal of Alloys and Compounds* **283**, 122–127 (1999).

24. Maisonneuve, V., Cajipe, V. B., Simon, A., Von Der Muhll, R. & Ravez, J. Ferrielectric ordering in lamellar CuInP 2 S 6. *Phys. Rev. B* **56**, 10860–10868 (1997).

25. Bagus, P. S., Ilton, E. S. & Nelin, C. J. The interpretation of XPS spectra: Insights into materials properties. *Surface Science Reports* **68**, 273–304 (2013).

26. Susner, M. A., Rao, R., Pelton, A. T., McLeod, M. V. & Maruyama, B. Temperature-dependent Raman scattering and x-ray diffraction study of phase transitions in layered multiferroic CuCrP2S6. *Physical Review Materials* **4**, 104003 (2020).

27. Selter, S. *et al.* Crystal growth, exfoliation, and magnetic properties of quaternary quasi-two-dimensional CuCrP2S6. *Physical Review Materials* **7**, 033402 (2023).

28. Biesinger, M. C. *et al.* Resolving surface chemical states in XPS analysis of first row transition metals, oxides and hydroxides: Cr, Mn, Fe, Co and Ni. *Applied Surface Science* **257**, 2717–2730 (2011).

29. Giannozzi, P. et al. Advanced capabilities for materials modelling with Quantum ESPRESSO. *Journal of Physics: Condensed Matter* **29**, 465901 (2017).

## Acknowledgements

This material is based upon work supported by the National Science Foundation (NSF) under Grant 1945364. Work done by X.L. was also supported by the U.S. Department of Energy (DOE), Office of Science, Basic Energy Science (BES) under Award DE-SC0021064. The XPS measurements are performed on a PHI Genesis system funded by the National Science Foundation under Grant No. 2216008. XRD measurements were supported by an NIH S11 OD028585 award for a Bruker D8 Venture Diffractometer, housed in the Chemical Instrumentation Center, Boston University.

## Author contributions

J.T. and X.L. conceived the idea and experimental plan. R.L. and J.A.B. performed temperature-dependent XPS measurements. J.T. performed the temperature dependent XRD measurements. J.T. performed crystal-field multiplet simulations. J.T. and X.L. wrote the manuscript with inputs from all other authors.

## Competing interests

The authors declare no competing interests.

**Supporting Information**

ΔSCF calculations; SEM/EDS and Raman characterization; room-temperature and temperature-dependent XPS analyses; CCPS XRD refinement results; charge compensation and valence-band spectra; Cu 2p fitting at 146 K; and low-temperature charge-density-difference calculations.

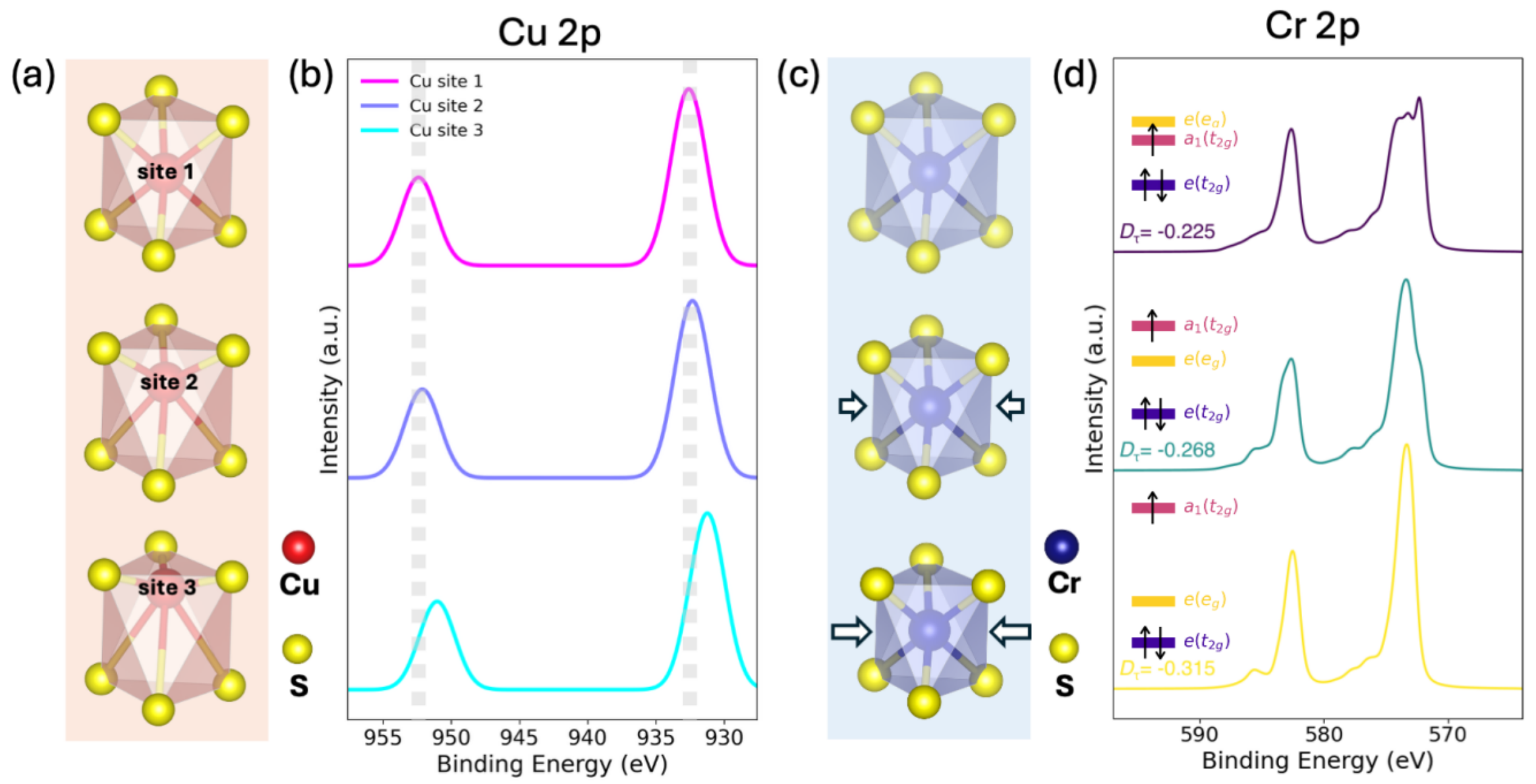

**Figure 1.** Theoretical modeling of core-level XPS evolution of Cu *2p* and Cr *2p* in the $CuS_6$ and $CrS_6$ octahedra under $Cu^+$ ion migration and lattice distortion. (a) Three representative Cu coordination environments (sites 1–3) produced by symmetry-lowering $Cu^+$ ion migration in $CuS_6$ octahedra. (b) Simulated Cu *2p* spectra with the BE shifts to lower binding energy from site 1 to site 3. (c) Reduced-symmetry $C_{3v}$ crystal-field model of the $CrS_6$ octahedra. From top to bottom, the lattice is more squeezed. (d) CFM simulations of the Cr *2p* spectra as a function of the trigonal distortion strength $D\tau$. Insets show diagonal Cr 3d crystal-field matrix elements with schematic spin arrows. $D\tau = -0.225$, $-0.268$, and $-0.315$ eV; these illustrative configurations do not specify the eigenstates of the full mixed Hamiltonian.

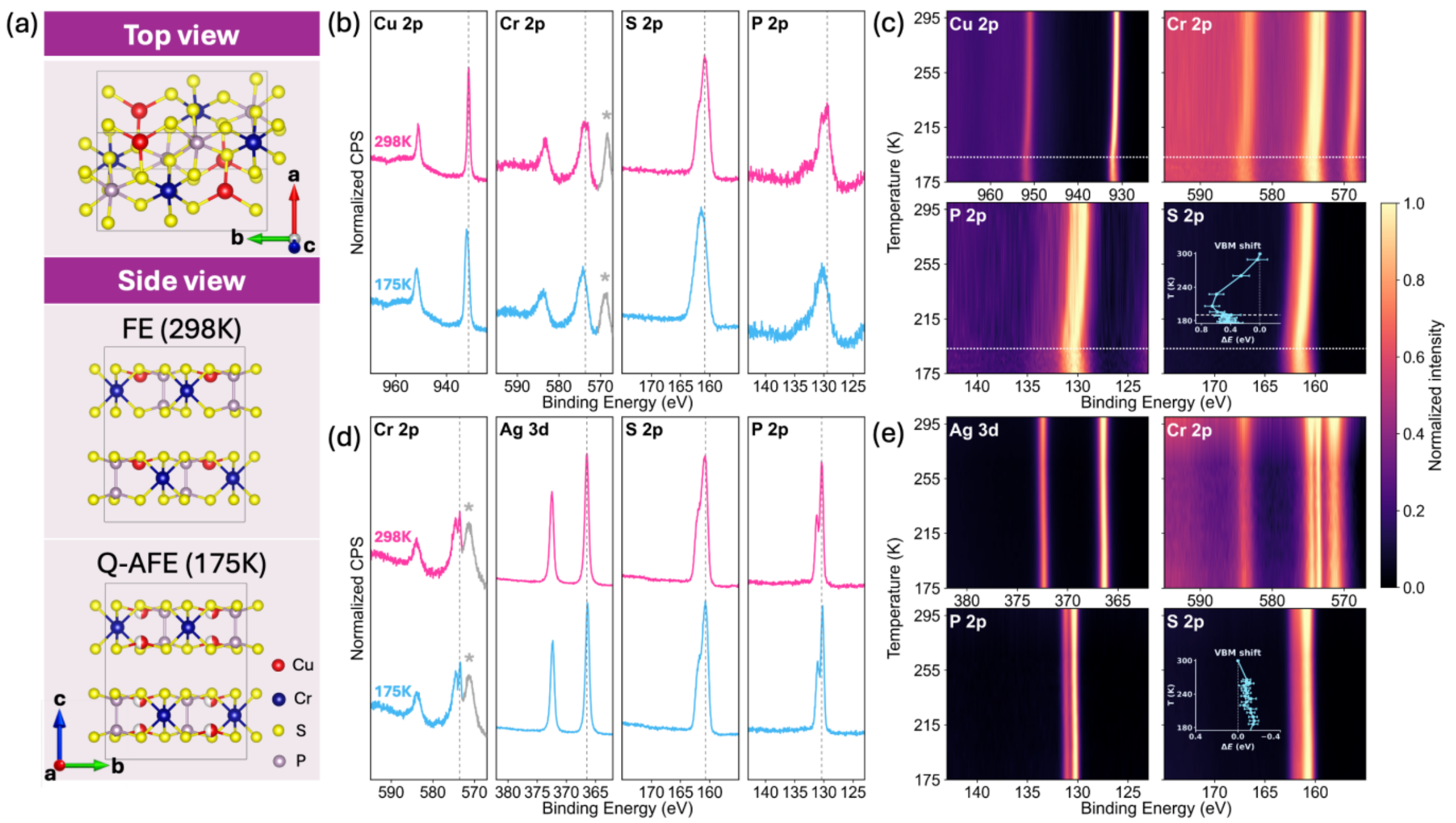

**Figure 2.** Temperature-dependent XPS across the ferroelectric phase transition in CCPS with ACPS as a reference. (a) Crystal structure of CCPS (top and side views) comparing the FE (298 K) and Q-AFE (175 K) states. (b) Cu *2p*, Cr *2p*, S 2p, and P *2p* spectra of CCPS at 175 and 298 K. Asterisks mark additional contributions (gray); low-binding-energy features in the Cr regions are provisionally assigned to Cu LMM in CCPS and Ag $3p_{3/2}$ in ACPS. (c) CCPS core-level heat maps; the white dashed line marks 190 K. Inset: VBM offset relative to 299.6 K and the white dash line represents 190K. (d) Corresponding ACPS core-level spectra at 175 and 298 K. (e) ACPS core-level heat maps. Inset: VBM offset relative to 300 K. VBM positions are obtained from the intersection of linear leading-edge and baseline fits; error bars show propagated regression standard errors.

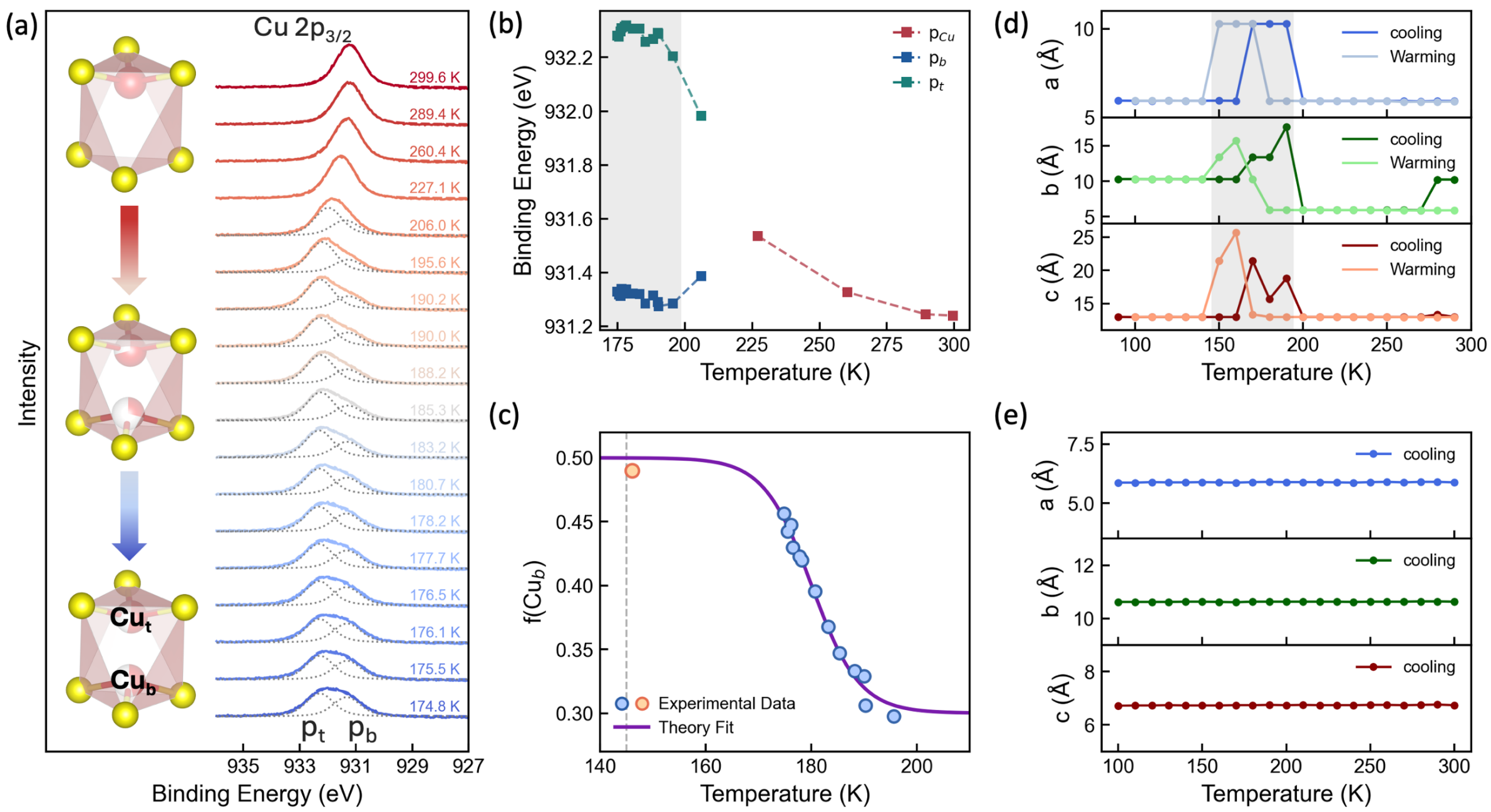

**Figure 3.** Cu *2p* spectral evolution revealing an intermediate state of ferroelectric ion migration. (a) Cu $2p_{3/2}$ core-level spectra of CCPS from 175 to 298 K. The inset illustrates gradual $Cu^+$ ion migration between the top and bottom sites. (b) Binding energy of Cu $2p_{3/2}$ versus temperature, showing peak splitting below ~206 K. The gray shaded region (below ~190 K) indicates the phase-transition regime. (c) Bottom-site population fraction $n_{Cu_b}$ versus temperature, fitted with a two-state Boltzmann sigmoid. (d, e) Lattice constants *a*, *b* and *c* from temperature-dependent single-crystal XRD for CCPS (d) and ACPS (e).

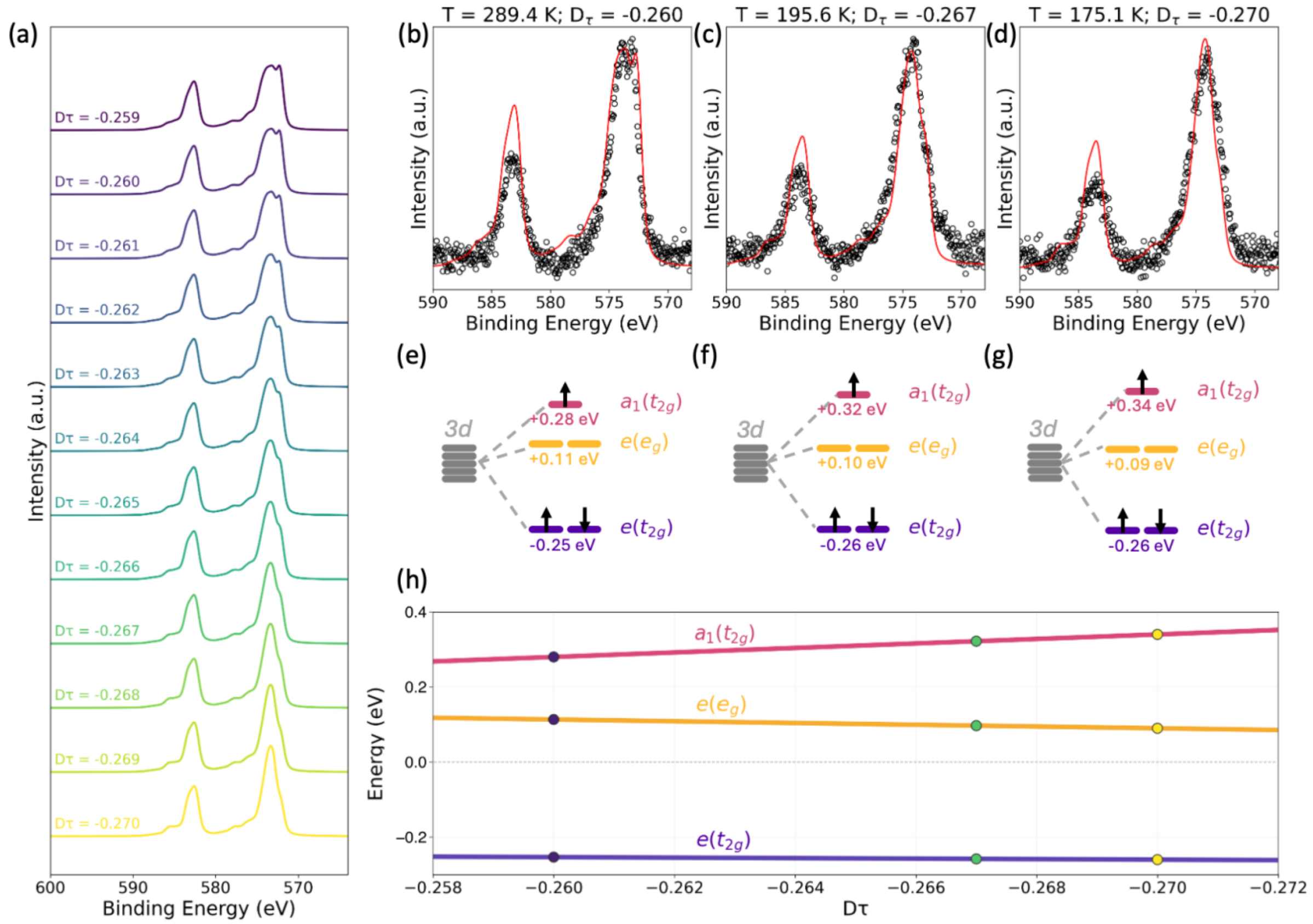

**Figure 4.** Local distortion-driven *d*-electron correlation revealed by Cr *2p* XPS. (a) CFM simulations of the Cr *2p* spectra for a range of trigonal distortion values Dτ from -0.270 to -0.259. (b–d) CFM simulations (red curve) compared with experimental Cr 2p spectra (black open circles) at 289.4 K, 195.6 K and 175.1 K, each compared with a single Dτ value without changing the formal Cr valence. (e–g) Diagrams of the diagonal Cr *3d* crystal-field matrix elements, showing $a_1(t_2g)$ rising away from $e(e_g)$ as Dτ becomes more negative. The three sets of colored dots correspond to the conditions in panels (b)–(d). (h) Diagonal crystal-field matrix elements for the $e(e_g)$ (yellow), $a_1(t_2g)$ (pink), and $e(t_2g)$ (purple) components plotted as a function of Dτ. The color-coded circles (purple, green, yellow) mark the three Dτ values shown in panels (b–g).

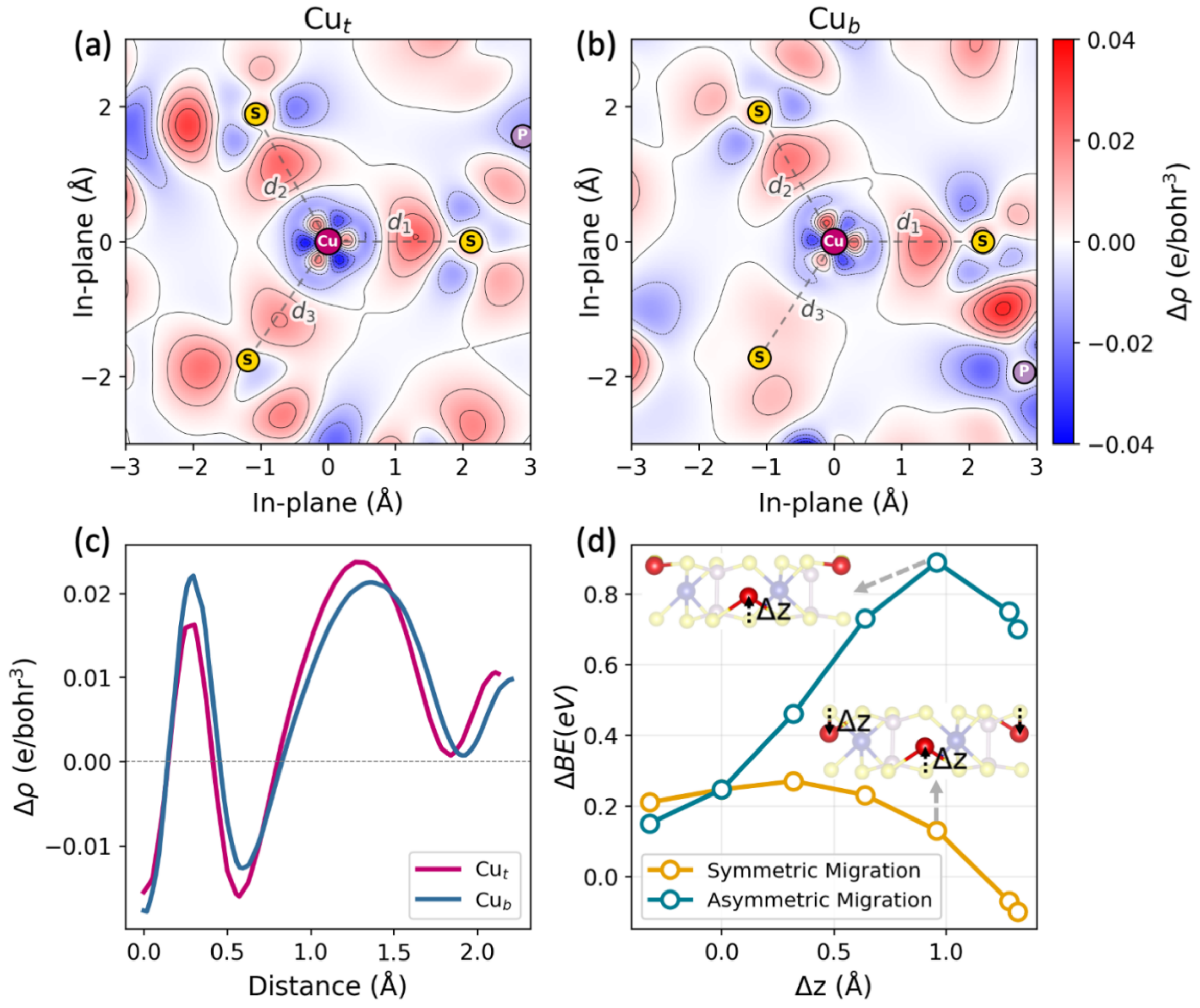

**Figure 5.** DFT-calculated charge-density difference and binding-energy response for the two Cu sites ($Cu_t$ and $Cu_b$). (a, b) Two-dimensional charge-density difference maps (Δρ) in in-plane cross-section through (a) $Cu_t$ and (b) $Cu_b$. Red and blue denote charge accumulation and depletion, respectively. Dashed lines mark the Cu–S distances, with $d_1$ = 2.13 Å, $d_2$ = 2.17 Å, and $d_3$ = 2.20 Å for $Cu_t$ (a) and $d_1$ = 2.21 Å, $d_2$ = 2.22 Å, and $d_3$ = 2.29 Å for $Cu_b$. (c) Line profiles of Δρ along the Cu–S direction ($d_1$) for $Cu_t$ and $Cu_b$; the dashed line marks Δρ = 0. (d) Calculated binding-energy difference (ΔBE) between $Cu_t$ and $Cu_b$ as a function of the $Cu^+$ ion migration along c-axis with the displacement Δz for symmetric (both Cu translated by Δz toward the Cr plane) and asymmetric (only $Cu_b$ translated) migration pathways.